\documentclass[sigconf]{acmart}
\renewcommand\footnotetextcopyrightpermission[1]{} 

\usepackage{diagbox}
\usepackage{hhline}
\usepackage{amsfonts} 
\usepackage{algorithmic}
\usepackage{textcomp}
\usepackage{url}
\usepackage[nolist]{acronym}
\usepackage{booktabs}
\usepackage{enumitem}
\usepackage{multirow}
\usepackage{float}
\usepackage{multirow}
\usepackage{cleveref}
\usepackage{makecell}
\usepackage{xcolor}
\usepackage{colortbl}

\begin{acronym}
\acro{SLR}[SLR]{systematic literature review}
\acro{SMR}[SMR]{state machine replication}
\acro{POW}[PoW]{proof-of-work}
\acro{POS}[PoS]{proof-of-stake}
\acro{DLT}[DLT]{distributed ledger technology}
\acro{MDP}[MDP]{Markov decision process}
\acro{BFT}[BFT]{Byzantine fault-tolerant}
\acro{PKI}[PKI]{public key infrastructure}
\end{acronym}

\begin{document}
\sloppy
\title{Unsealing the secrets of blockchain consensus: A systematic comparison of the formal security of proof-of-work and proof-of-stake}
  
\renewcommand{\shorttitle}{Systematic comparison of the formal security of proof-of-work and proof-of-stake}

\author{Iv\'{a}n Abell\'{a}n \'{A}lvarez}\orcid{0000-0003-4670-433X}
\orcid{0000-0003-4670-433X}
\affiliation{%
  \institution{Interdisciplinary Centre for Security, Reliability and Trust, University of }
  \country{Luxembourg} 
}
\email{ivan.abellan@uni.lu}

\author{Vincent Gramlich}\orcid{0000-0003-3070-1433}
\orcid{0000-0003-3070-1433}
\affiliation{%
  \institution{Branch Business \& Information Systems Engineering, Fraunhofer FIT}
  \country{Germany}
}
\email{vincent.gramlich@fit.fraunhofer.de}

\author{Johannes Sedlmeir}\orcid{0000-0003-2631-8749}
\orcid{0000-0003-2631-8749}
\affiliation{%
  \institution{Interdisciplinary Centre for Security, Reliability and Trust, University of }
  \country{Luxembourg} 
}
\email{johannes.sedlmeir@uni.lu}

\begin{abstract}
 With the increasing adoption of decentralized information systems based on a variety of permissionless blockchain networks, the choice of consensus mechanism is at the core of many controversial discussions. Ethereum's recent transition from~\ac{POW} to~\ac{POS}-based consensus has further fueled the debate on which mechanism is more favorable. While the aspects of energy consumption and degree of (de-)centralization are often emphasized in the public discourse, seminal research has also shed light on the formal security aspects of both approaches individually. However, related work has not yet comprehensively structured the knowledge about the security properties of \ac{POW} and \ac{POS}. Rather, it has focused on in-depth analyses of specific protocols or high-level comparative reviews covering a broad range of consensus mechanisms. To fill this gap and unravel the commonalities and discrepancies between the formal security properties of \ac{POW}- and \ac{POS}-based consensus, we conduct a systematic literature review over 26 research articles. Our findings indicate that~\ac{POW}-based consensus with the longest chain rule provides the strongest formal security guarantees. Nonetheless, \ac{POS} can achieve similar guarantees when addressing its more pronounced tradeoff between safety and liveness through hybrid approaches.
\end{abstract}

%
%

\ccsdesc[300]{Security and privacy~Formal security models}
\ccsdesc[300]{Security and privacy~Distributed systems security}

\keywords{Distributed ledger technology, dynamic availability, finality, liveness, PoS, PoW, safety}

\maketitle

\acresetall

\section{Introduction}\label{sec:intro}
The proliferation of cryptocurrencies and decentralized applications following the introduction of Bitcoin~\cite{nakamoto_bitcoin_2009} has spurred the need for effective designs of blockchain infrastructures with open (``permissionless'') participation. Blockchain technology provides a foundational data structure that enables the secure and intermediary-free transmission of both information and value in such decentralized networks~\cite{lacity2022blockchain}. At its core, a blockchain is a replicated, (probabilistically) immutable, ever-growing event log file that records all participants' transactions in a well-defined total order~\cite{nakamoto_bitcoin_2009, li_analysis_2019, butijn_blockchains_2020}. Transactions are ordered and assigned to sequential batches known as blocks, each of which is cryptographically linked to the preceding block to achieve tamper evidence. This append-only structure facilitates an efficient synchronization process. Corresponding agreement rules that make synchronization robust and provide a shared and consistent view of the database even in the presence of partial system outages and malicious activities by some participants are called ``consensus mechanisms''~\cite{garay_bitcoin_2015, graf_security_2021, wang_byzantine_2022}. 

The consideration of faulty nodes that may crash or act maliciously in permissionless networks naturally involves the consideration of Sybil attacks: the ability of an entity to subvert and solely control many bogus identities at negligible costs that may participate in the blockchain network and, thus, influence the outcome of agreement processes~\cite{douceur2002sybil,platt_sybil_2023}. In general, consensus mechanisms for permissionless blockchains hence combine decision rules with voting procedures and Sybil resistance mechanisms that linearly couple voting weight to a scarce resource~\cite{sedlmeir2020energy}. Bitcoin introduced a novel combination of a cryptographic data structure, Sybil resistance mechanism, economic incentives for participation, and agreement rules to implement such a blockchain network functioning in an open and decentralized setting~\cite{nakamoto_bitcoin_2009}. More specifically, provably invested computational power dictates a node's voting weight in its \ac{POW}-based longest chain consensus. However, while offering convincing heuristics and examples for security conditions like the maximum adversarial tolerance threshold for computational power acceptable for ensuring consistency (e.g., to prevent double-spending with high probability), the original Bitcoin paper did not derive formal security guarantees~\cite{nakamoto_bitcoin_2009}. Similar observations hold for the originally proposed~\ac{POS} in Peercoin~\cite{king_ppcoin_2012}, which achieves Sybil resistance by coupling a node's voting weight to its cryptocurrency coin holdings. 

Contrary to permissionless blockchains, there are also ``permissioned'' constructions in which the number and identity of nodes are specified at the start of the protocol and that involve well-defined processes to add or remove nodes during operation. Corresponding deterministic consensus mechanisms that tolerate a maximum number of faulty nodes are also known as~\ac{BFT} protocols and have been well-studied since the 1980s~\cite{lewispye2023byzantine}. Permissionless blockchains can be considered a strict generalization of permissioned constructions, with a considerably lower degree of control over participants and their behavior, which brought new opportunities and challenges to distributed systems designs~\cite{nakamoto_bitcoin_2009,wood2014ethereum}. Many established approaches to formalization and formal security analyses for permissioned networks turned out not to adequately accommodate these novel, naturally non-deterministic constructions. Consequently, alternative models and notions on the ideal functionality of distributed systems~\cite{graf_security_2021} have been proposed, alongside novel formal security properties permissionless blockchains should meet~\cite{garay_bitcoin_2015}. 

A significant challenge for organizations is the effective selection and management of permissionless blockchain-based infrastructures, which is manifested in discussions around corresponding design principles~\cite{risius_blockchain_2017,sai2021taxonomy}. The choice of the consensus protocol plays a crucial role. This role manifests in an ongoing controversial debate between the arguably two most prominent designs, namely~\ac{POW} and~\ac{POS}~\cite{guo_survey_2022, lewispye2023byzantine}. Permissionless blockchains are restricted by several impossibility results (see \Cref{sec:2-historical} for details), such as the incompatibility of finality (safety) and dynamic availability (liveness) under non-synchronous network conditions. Many design choices have a significant impact on the consensus protocol's security properties~\cite{pye_resource_2020,sankagiri_blockchain_2021,neu_availability-accountability_2022,buterin_incentives_2019, ramezan_analysis_2020}. In general, \ac{POW} is often touted for its security~\cite{gervais_security_2016,ouyang_pow_2021} but faces significant criticism due to its high energy consumption~\cite{sedlmeir2020energy}. This debate has witnessed significant contention particularly since the Ethereum blockchain introduced the ``Merge'', which marked its transition from~\ac{POW} to~\ac{POS} in September~2022. In particular, while \ac{POS}-based consensus undisputedly improves energy consumption substantially~\cite{rieger2022debate,CCRI_2022_Ethereum}, its impact on security is controversial~\cite{neu_availability-accountability_2022,neu_two_2022,wang_byzantine_2022}. However, while there is broad agreement that blockchain security aspects are crucial for ensuring reliability and trustworthiness~\cite[e.g.,][]{lewispye2023byzantine, lewispye2021,garay_bitcoin_2015, graf_security_2021, gazi_tight_2020, dembo_everything_2020, neu_availability-accountability_2022}, security often seems overlooked.

We observe two main dimensions of the prevailing academic discussion on the security aspects of~\ac{POW} and~\ac{POS} in permissionless blockchain designs. The first dimension is a dedicated stream in the economic literature that considers the degree of decentralization and the characteristics of the time evolution of the distribution of the scarce resource that impacts an entity's voting weight in consensus: Computational power (``hash rate'') in~\ac{POW} and capital (``staked crypto-assets'') in~\ac{POS}~\cite[e.g.,][]{arnosti2022bitcoin,rocsu2021evolution}. The second dimension, the focus of this work, represents formal security properties, many of which are based on a given distribution of voting weights and threshold assumptions on honest protocol execution. 
Some related studies already offer a high-level comparative analysis of characteristics of permissioned and permissionless consensus protocols, including security aspects~\citep[e.g.,][]{guo_survey_2022,xiao_survey_2020}. Two recent works~\cite{yadav_comparative_2023, singh_survey_2022} provide a comprehensive survey of various consensus mechanisms and corresponding key components, including \ac{POW} and \ac{POS}-based Sybil resistance mechanisms and permissioned consensus approaches. \cite{singh_survey_2022} further extends and categorizes the literature while highlighting the main advantages, limitations, and applicability of various proof-of-x mechanisms. However, these works do not focus on comparing the formal security aspects related to computational limitations, incompatibilities, known vulnerabilities, and associated mitigation strategies, and thus do not provide an in-depth understanding of the nuanced differences between~\ac{POW} and~\ac{POS}. On the other hand, seminal research on the formal security properties of permissionless consensus has primarily focused on analyzing individual blockchains and their consensus mechanisms, e.g., for~\ac{POW}~\cite{garay_bitcoin_2015,gazi_tight_2020} and~\ac{POS}~\cite{dembo_everything_2020,nakamura_refinement_2019}. On the other hand, \cite{graf_security_2021} contributes a formal, consensus-agnostic framework to model protocols for security analysis, and~\cite{gazi_tight_2020} provides an analytical approach to describing the security properties of a~\ac{POW} consensus protocol that is readily generalizable to cover also~\ac{POS}-based constructions. However, there seems to be no general agreement on the appropriate collection of core security properties for permissionless consensus.
In particular, a comprehensive and detailed comparison between prominent design choices for permissionless consensus mechanisms, and in particular between popular instantiations of~\ac{POW} and~\ac{POS}, is lacking. This paper closes this research gap by systematically analyzing the commonalities and differences between formalization aspects and formal security properties to comprehensively capture and compare the trade-offs inherent to \ac{POW} and \ac{POS}-based consensus mechanisms for permissionless blockchains. We hence ask the following two research questions:

\textbf{RQ1.} \textit{What are commonly considered security properties for consensus mechanisms in permissionless blockchains?}\label{rq1}

\hspace{-0.075cm}\textbf{ RQ2.} \textit{What are the commonalities and differences between~\ac{POW}- and~\ac{POS}-based consensus mechanisms regarding those formal security properties?}\label{rq2}

To answer these research questions, we ground our study in peer-reviewed journals and conferences, following established guidelines for conducting \acp{SLR}~\cite{kitchenham_procedures_2004, brereton_lessons_2007}.
 
\section{Background}
\subsection{Historical Overview of Decentralized Systems Security}\label{sec:2-historical}

To understand the functionalities and security characteristics of information systems, research often employs formalization that takes into account the behavior of the system's individual components. A common formalization model for blockchains is~\ac{SMR}, which characterizes the behavior of every individual node of a distributed system~\cite{schenider_90_smr}.
\Ac{SMR} thus provides an abstract model for a system of deterministic machines (``nodes'') that handle information processing by individual storage and mutual communication~\cite{schenider_90_smr}. \Ac{SMR} captures an ordered sequence of inputs (``transactions''), with the goal of ensuring a consistent and logical execution such that from the perspective of clients, the decentralized and in particular distributed system consisting of many nodes behaves like a highly reliable centralized system (e.g., running on a faultless server).~\Ac{SMR} is widely used to formalize and prove the security of permissioned consensus protocols~\cite{schenider_90_smr}. The consensus protocol accordingly updates the current state of each local~\ac{SMR} node~\cite{schenider_90_smr}.~\Ac{SMR} formally ensures consensus protocol correctness (i.e., qualifies as \ac{BFT}) if a system satisfies the following key properties~\cite{lamport_proving_1977}: 1)~\emph{Safety}: The protocol does not produce contradictory states among non-faulty nodes, thus ensuring a consistent and persistent view~\cite{garay_bitcoin_2015, lewispye2023byzantine}; and 2)~\emph{Liveness}: correct processing of transactions will eventually happen upon correct input~\cite{owicki_proving_1982, alpern1987recognizing}, i.e., the distributed system keeps making meaningful progress over time and is useful to interact with~\cite{lewispye2021}. Whether or not a distributed system satisfies the safety and liveness properties can vary depending on the number of faulty nodes, the use of additional building blocks (e.g., digital signatures and public key infrastructure for authenticated messaging~\cite{dwork_gst_88,guerraoui_consensus_2019}), and other conditions, such as network reliability and performance~\cite{gervais_security_2016,lamport2019time}. For representing the mentioned network properties, three models are widely used, namely synchrony, partial synchrony, and asynchrony. A synchronized network communicates reliably with a bounded maximum transmission time~\cite{lewispye2021,simons_90_clocksyn}.
Asynchrony, on the other hand, considers potentially unbounded communication time delays and therefore offers no control on the number of lost messages~\cite{lewispye2021,simons_90_clocksyn}. Partial synchrony is between these two extremes, denoting two synchronization states: Periods of asynchrony that last for an unknown time, and periods of synchrony that eventually occur~\cite{lewispye2021, dwork_gst_88}. Partial synchrony is widely considered to be an appropriate model for the Web~\cite{gilbert_lynch_web_02}.

The Dolev-Strong protocol presented in 1983~\cite{ds83} marks one of the first solutions for permissioned consensus under synchrony. By making use of authenticated messaging, it can handle any number of faulty nodes, as digital signatures on all messages ensure that malicious nodes that supply different nodes contradictory information are detected. The protocol involves honest nodes incrementally affirming a decision by adding digital signatures, resulting in a~\ac{BFT} solution. However, distributed systems with potentially faulty nodes are fundamentally constrained when facing unreliable network conditions. In~1985,~\cite{fischer_impossibility_1985} proved the FLP~theorem that states that in a permissioned system under asynchronous network assumptions, no deterministic solution for consensus exists even if only a single node may crash. However, solutions that achieve both safety and liveness with high probability exist when non-deterministic components are used. HoneyBadgerBFT~\cite{miller_honey_2016}, proposed in 2016, is an example of such a non-deterministic permissioned consensus protocol with practical performance under asynchronous network conditions. Furthermore, the CAP~theorem dictates fundamental security principles for distributed systems under stronger synchronicity assumptions. After the initial conjecture by Brewer in 2000~\cite{brewer_cap_theorem}, the CAP theorem was formalized and proved in 2002~\cite{gilbert_lynch_web_02}. The CAP acronym stands for \emph{Consistency} (i.e., safety in the sense that all operations execute in the same order for every available honest node), \emph{Availability} (i.e., liveness on messages that are eventually delivered, thus implying that the execution terminates), and \emph{Partition tolerance} (i.e., tolerating certain network failure events where lost messages lead to a split of the network into isolated parts)~\cite{brewer_cap_theorem,gilbert_lynch_web_02}. The theorem states that distributed systems can only fulfill only two out of these three properties in the partially synchronous (and, therefore, also in the asynchronous) setting~\cite{gilbert_lynch_web_02}. Yet, solutions exist in partial synchrony if the consistency requirement is weakened~\cite{gilbert_lynch_web_02, fischer_impossibility_1985}.

While some recent advances have been made in permissioned consensus, e.g., substantial performance improvements by reductions of message complexity in non-deterministic~\ac{BFT} protocols for asynchronous networks~\cite{miller_honey_2016} or deterministic~\ac{BFT} protocols under partial synchrony~\cite{abraham2018hot,neu_ebb-and-flow_2021}, research in the last decade has primarily focused on understanding the emerging permissionless systems. It turns out that in the permissionless setting, there is an analogous impossibility result to the CAP theorem~\cite{sankagiri_blockchain_2021, pye_resource_2020}. It shows fundamental incompatibilities between finality and dynamic availability properties under a partially synchronous network. This incompatibility is an essential characteristic that distinguishes \ac{POW}- and \ac{POS}-based consensus protocols~\cite{lewispye2021,lewispye2023byzantine} (see~\Cref{sec:formal}). \textit{Finality} defines the consistency of states and irreversibility of transactions or blocks. While finality is naturally satisfied in deterministic \ac{BFT} protocols, many non-deterministic consensus protocols only provide \textit{probabilistic finality}, e.g., with the probability of reversibility decreasing exponentially with each new block added to the blockchain in longest-chain rule~\ac{POW}~\cite{pye_resource_2020,garay_bitcoin_2015} (see \Cref{sec:pow}). \textit{Dynamic availability}, on the other hand, refers to a network characteristic that can provide liveness even in the presence of arbitrary fluctuations in the participating nodes or their voting power~\cite{pye_resource_2020}. 
Intuitively, the reason behind the dynamic availability-finality dilemma is the fact that from the perspective of an honest node, network partitions are indistinguishable from diminishing network participation~\cite{pye_resource_2020}. Consequently, dynamically available networks must ``keep growing the chain'' even in the case of a network partition~\cite{sankagiri_blockchain_2021}, which may produce a split view~\cite{sankagiri_blockchain_2021}. Such a split view, by definition, compromises consistency (safety) if both sides are finalizing transactions~\cite{pye_resource_2020}.

\subsection{Proof-of-Work and Proof-of-Stake Constructions}\label{sec:2-block}
Permissionless settings need to ensure the consistency of blockchain nodes among honest participants under an honest-majority assumption (in some metric). Consequently, they must account for faulty nodes and in particular for Sybil attacks (see~\Cref{sec:intro}). Permissionless blockchain consensus hence relies on the integration of a Sybil resistance mechanism in addition to a rule for choosing the valid state among potentially multiple options (``forks''). These mechanisms require the expenditure or investment of a scarce resource for participation in consensus~\cite{gervais_security_2016, thomsen_formalizing_2021}.~\Ac{POW} and~\ac{POS} are Sybil resistance mechanisms that also incentivize honest behavior in consensus through rewards. Honest behavior is typically rewarded in the form of fixed block rewards and variable transaction fees for block producers, as well as compensation for further activities in consensus. These incentives are distributed as tokens of the native cryptocurrency that every permissionless blockchain network needs as the basis for its compensation mechanism. Adversarial attempts are typically penalized by depriving rewards for the expenditure of the scarce resource or corresponding opportunity costs, as well as potential capital forfeiture in~\ac{POS}~\cite{neu_availability-accountability_2022,nakamura_refinement_2019}.

The first ever introduced permissionless blockchain, Bitcoin, is governed by~\ac{POW} in combination with the longest-chain rule (``Nakamoto consensus'')~\cite{nakamoto_bitcoin_2009}.~\Ac{POW} defines a verifiably computationally intensive mechanism that on average proves the provisioning of a certain amount of computing hardware and electric energy~\cite{dwork_pow_92} to regulate the block production rate (i.e., block proposers and the time intervals at which blocks are appended to the global ledger). Blocks represent a solution to a cryptographic puzzle that is used to determine the average amount of computational power provided by a participant. Solving this puzzle makes the node operators (``miners'') eligible to append the block under consideration to the existing chain. The canonical longest chain is defined to be the chain that requires the largest computational effort for construction, i.e., ``length'' is determined according to a weighted sum over all blocks, where a block's weight is determined by the difficulty of the puzzle solved by the block.

Permissionless blockchain transaction throughput rate is well-known to be low: Throughput defines the computational, bandwidth, and storage resource requirements of each node; such that only a low throughput keeps the barrier to participation sufficiently moderate to make decentralization in the long run possible~\cite{sai2021taxonomy}. Moreover, low throughput usually involves small blocks (i.e., fast propagation) with a slow block production rate, which is beneficial for security: It decreases the probability of unintentional forking, where honest nodes try to build on a block that is not the most recent one because they have not yet received the latest block~\cite{garay_bitcoin_2015, gervais_security_2016}. However, this low throughput of blockchains compared to centralized systems restricts the scalability~\cite{ghost_15} and is hardly compatible with the real-time requirements in organizations~\cite{guggenberger2022fabric}. A faster block production can help both to reduce transaction confirmation latencies and achieve slightly higher transaction throughput by facilitating a more continuous use of computation and bandwidth resources as long as storage is not the bottleneck.
To increase the block production rate without compromising security, the Greedy Heaviest Observed Subtree (GHOST) protocol was proposed~\cite{ghost_15} for permissionless blockchain consensus protocols. In GHOST, new blocks are appended to the previous most voted block (i.e., higher weight) while stale but valid blocks still influence the chain~\cite{nakamura_refinement_2019, neu_two_2022, neu_ebb-and-flow_2021, wang_byzantine_2022}. The canonical chain is, therefore, the heaviest in terms of votes in contrast to the longest in terms of counting or weighted by the difficulty of the~\ac{POW} puzzle~\cite{neu_two_2022}. These constructions by which honest nodes decide where to append a new block they propose are often referred to as ``fork-choice rules''.

In~\ac{POW}, miners face direct costs for participation in consensus through their contribution of hardware and electricity, which puts a strong incentive for them to behave honestly: The cryptographic puzzle is dependent on a specific batch of transactions and a previous block, i.e., the chain a miner decides to extend~\cite{garay_bitcoin_2015}. Consequently, miners have to wisely choose how to use their resources, and if they use them on a block deemed invalid by other miners, or for extending a chain that is not currently the longest one, they face a substantial risk that their block will not be respected by the majority of other miners, i.e., they gain no rewards and their resources are wasted~\cite{gervais_security_2016}. \ac{POS} replaces hardware and electric energy as a scarce resource to which voting power is coupled by capital in the form of ownership of native cryptocurrency tokens, which is also publicly verifiable through the transparent accounting in the permissionless blockchain network~\cite{buterin_incentives_2019}. Participants deposit capital (``stake'') to signal their willingness to participate in consensus. They are then selected to act as block proposers or to validate and attest to blocks as part of a committee, typically with probability equal or close to their share of the total stake, using some source of (pseudo-) randomness generated in the protocol~\cite{kiayias_ouroboros_2017, buterin_incentives_2019}. 

Consequently, contrary to \ac{POW}, \ac{POS} faces the issue of ``costless simulation'', leading to the \textit{nothing-at-stake} problem: Any node can at negligible cost create different blocks at the same height (``equivocation'') that could potentially be both deemed correct and included by honest nodes~\cite{nakamura_refinement_2019}. In the case of the existence of forks, i.e., alternative chains with equal height, nodes even have an incentive to engage in equivocation because this makes them eligible for rewards in any future scenario chosen by the majority of nodes. Hence, the costless simulation issue would imply that a fork may never be resolved. To address this shortcoming, miners can be held accountable for observed misconduct, including equivocation, through their staked capital (``slashing'')~\cite{buterin_incentives_2019, nakamura_refinement_2019, neu_availability-accountability_2022}. Another way to address costless simulation and the risks of equivocation is to achieve immediate finality by using the Sybil resistance mechanism only to determine a subset of nodes that subsequently run a (permissioned) \ac{BFT} protocol~\cite{sankagiri_blockchain_2021,rana_optimal_2022}. This approach and check-pointing services that continuously mark a set of blocks as finalized after a relatively short time can also effectively prevent long-range attacks in which attackers have already disposed of their collateral at the time of launching an attack with an alternative chain~\cite{sankagiri_blockchain_2021}. Hence, regardless of the fork-choice rule, many common constructions of \ac{POS} systems employ permissioned~\ac{BFT} protocols to guarantee safety, either for immediate finalization or check-pointing services~\cite{buterin_incentives_2019,neu_ebb-and-flow_2021, sankagiri_blockchain_2021}. Alternatively, modifications to the fork-choice rule can yield secure constructions if the leader election process is sound~\cite{kiayias_ouroboros_2017}. The current state of the Ethereum consensus protocol combines the GHOST protocol with Casper the Friendly Finality Gadget~(FFG) (this combination is often called ``Gasper'')~\cite{neu_ebb-and-flow_2021, wang_byzantine_2022}. Casper~(FFG) combines a~\ac{POS}-based fork choice rule, weighted by attached attestations' stake, and a~\ac{BFT}-style protocol that provides a check-pointing service for finalizing blocks~\cite{buterin_incentives_2019, neu_availability-accountability_2022}. The eligibility of block proposers and validators is drawn from the required deposited stake from which smaller-sized committees are formed every round~\cite{buterin_incentives_2019, neu_ebb-and-flow_2021} or randomly selected by weighted stake~\citep{kiayias_ouroboros_2017} to validate blocks. 
\section{Method}
To answer our research questions (see \Cref{sec:intro}), we comprehensively collected relevant academic works by conducting an~\ac{SLR} following the guidelines of~\citet{kitchenham_procedures_2004}. Based on a basket of literature we had already collected and investigated in a preliminary study, we defined our search string:
\textit{("security analysis" OR "adversarial attacks" OR liveness OR finality OR safety) AND (proof*of*stake OR proof*of*work OR Nakamoto OR Bitcoin OR Ethereum) AND (protocol OR consensus)}. 
We then applied this search string to a set of academic databases based on their relevance to the study topic~\cite{kitchenham_procedures_2004, kitchenham_systematic_2009}.
We selected four popular computer science databases that encompass journals and conference proceedings: ACM~Computing~Library, IEEE~Xplore, ScienceDirect, and SpringerLink. For ACM~Digital~Library and IEEE~Xplore, we applied the search string as specified. To keep the effort for searches manageable, we further tailored the search string for ScienceDirect and SpringerLink: We excluded the keywords \textit{liveness}, \textit{finality}, and \textit{safety} to reduce the otherwise n\,=\,1763 results in SpringerLink down to~558. Due to limitations in the number of Boolean operators supported in ScienceDirect (max.~8), we ran two independent queries, including either ``\textit{protocol}'' or ``\textit{consensus}''. These two queries yielded 38~and 39~papers, respectively. Through building the union of these results (i.e., removing duplicates), we arrived at an initial selection of 41~publications from ScienceDirect.

\begin{figure}[ht]
\centering
\includegraphics[width=0.75\columnwidth,trim=1.3cm 1.3cm 1.3cm 1.3cm, clip]{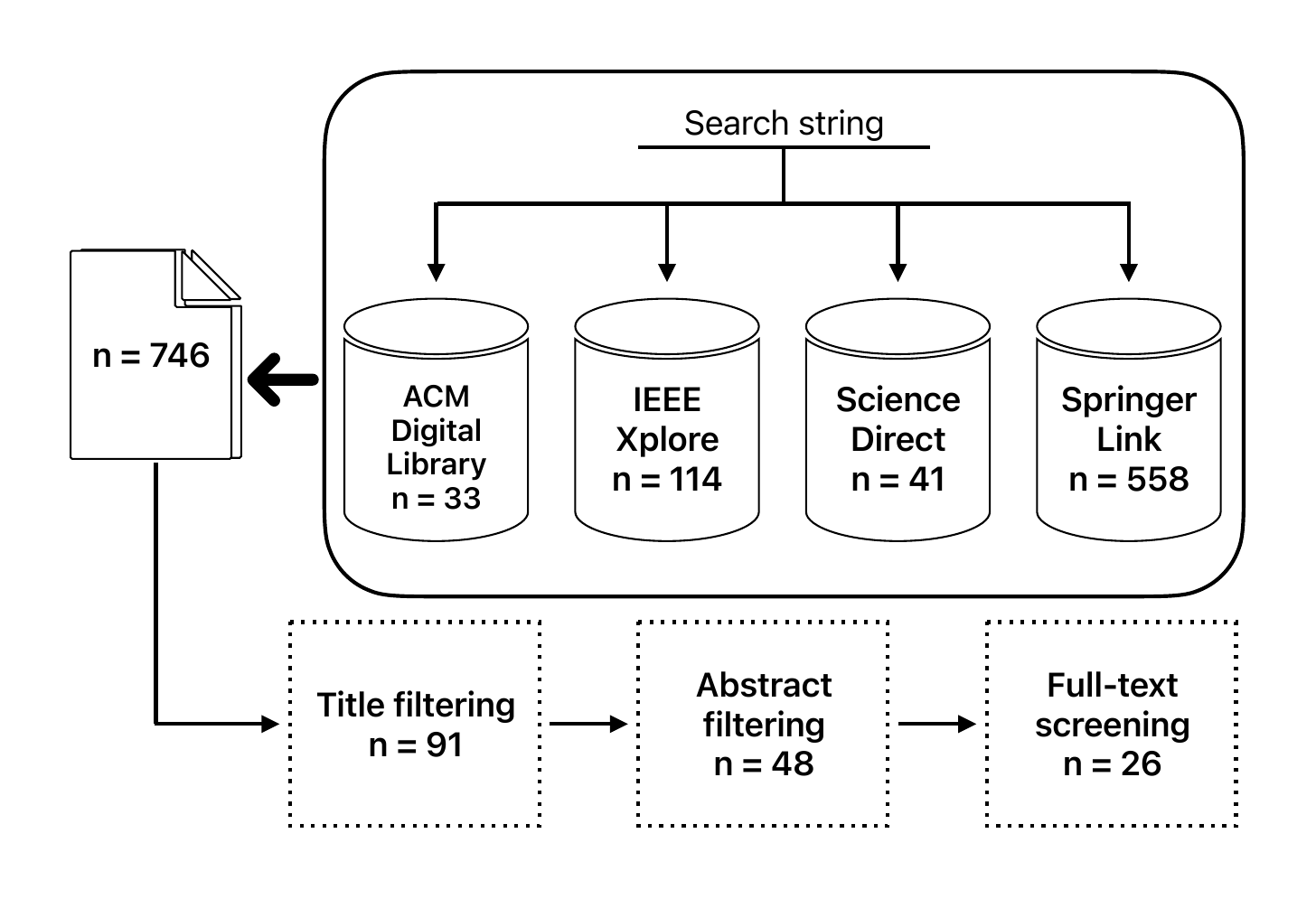}
    \caption{Systematic literature review process.}
    \label{fig:process}
\end{figure}

We illustrate the subsequent paper selection process in \Cref{fig:process}. To guide the \ac{SLR}, we defined a set of inclusion and exclusion criteria that we applied to the screening and filtering of title, abstract, and full-text~\cite{kitchenham_procedures_2004, kitchenham_systematic_2009,brereton_lessons_2007}. 
We include works formalizing blockchain characteristics and properties, e.g., by describing an ideal functionality, analyses of consensus protocols that involve \ac{POW}- and~\ac{POS}-based Sybil resistance mechanisms, investigations of corresponding security implications, attacks against such systems, and corresponding mitigation approaches. On the other hand, we exclude publications that 1)~put an exclusive focus on permissioned consensus or Sybil resistance mechanisms beyond \ac{POW} and \ac{POS}, 2)~consider only alternative blockchain constructions unrelated to \ac{POW} or \ac{POS}-based consensus mechanism, 3)~represent high-level surveys on a broad set of consensus mechanisms, or 4)~lack a formal evaluation method of the prescribed model. 
We carried out the filtering process by applying the inclusion and exclusion criteria at every step. Initially, our search string yielded a total of 746~results across all four databases. A subsequent title screening resulted in 91~remaining papers. We then evaluated these publications' abstracts and removed duplicates, narrowing our selection down to 48~publications. Lastly, a full-text analysis of these papers yielded our final selection of~26 publications. 

We classified the selected literature according to the topics it addresses among three groups: blockchain formalization,~\ac{POW}-based blockchain constructions, and~\ac{POS}-based ones. We then extracted the security properties discussed in these groups and mapped them into related groups to reflect, for instance, the close connection between safety, consistency, and finality; as well as the tight relationship between liveness, dynamic availability, chain quality, and chain growth (see \Cref{sec:formal}). We thus formed a structured overview of blockchain security properties that we use as a basis for answering~RQ2.

\section{Results}
The following section presents the results we extracted from the selected literature. In \Cref{tab:1}, we categorize all the relevant literature with respect to the consensus protocol constructions it analyzes and the considered security properties. \Cref{tab:1} represents the possible construction of consensus protocols via combinations of Sybil resistance mechanisms and a fork-choice rule according to what we have defined in our literature selection criteria, namely~\ac{POW} and~\ac{POS}. It also features the relation of all papers with the identified security properties.~\Cref{sec:formal} discusses abstract models of the functionalities and the security properties of permissionless blockchains. We then provide a comprehensive overview of formal security properties of~\ac{POW} and \ac{POS}-based consensus mechanisms in \Cref{sec:pow} and \Cref{sec:pos}, respectively. 

\begin{table*}
\centering
\caption{Classification of the literature selected in our \ac{SLR}.}
\label{tab:1}
\resizebox{\textwidth}{!}{%
\begin{tabular}{|l|c|c|c|c|c|c|c|c|} 
\arrayrulecolor{black}\hhline{~|--------}
\multicolumn{1}{c|}{}& \multicolumn{8}{c|}{{\cellcolor{gray!60}}\large\textbf{Consensus protocol}}\\

\arrayrulecolor{black}\hhline{~|--------}
\multicolumn{1}{c|}{}& \multicolumn{7}{c|}{{\cellcolor{gray!15}}\textbf{Sybil-resistant mechanism}}&\cellcolor{gray!15}\\ 

\arrayrulecolor{black}\hhline{~|------->{\arrayrulecolor{gray!15}}->{\arrayrulecolor{black}}|}\arrayrulecolor{black}
\multicolumn{1}{c|}{}& \multicolumn{3}{c}{{\cellcolor{gray!15}}\textbf{Proof-of-Work}}&
\multicolumn{3}{|c|}{{\cellcolor{gray!15}}\textbf{Proof-of-Stake}} & \cellcolor{gray!15} &\cellcolor{gray!15}\\ 

\arrayrulecolor{black}\hhline{~|------>{\arrayrulecolor{gray!15}}->{\arrayrulecolor{black}}|>{\arrayrulecolor{gray!15}}->{\arrayrulecolor{black}}|}\arrayrulecolor{black}
\multicolumn{1}{c|}{}& \cellcolor{gray!15} & \multicolumn{4}{c|}{\cellcolor{gray!15}\textbf{Longest-chain}}& \cellcolor{gray!15}& \cellcolor{gray!15}& \cellcolor{gray!15}\\

\arrayrulecolor{black}\hhline{-|>{\arrayrulecolor{gray!15}}->{\arrayrulecolor{black}}|---->{\arrayrulecolor{gray!15}}->{\arrayrulecolor{black}}|>{\arrayrulecolor{gray!15}}->{\arrayrulecolor{black}}|>{\arrayrulecolor{gray!15}}->{\arrayrulecolor{black}}|}\arrayrulecolor{black}
{\cellcolor{gray!60}}\diagbox{\large\textbf{Security properties}}{\large\textbf{Fork-choice rule}~} &\multirow{-2}{*}{\cellcolor{gray!15}\textbf{GHOST}}& \cellcolor{gray!15} & \multicolumn{2}{c|}{\cellcolor{gray!15}\textbf{BFT/Quorum-checkpointed}} & \cellcolor{gray!15} & \multirow{-2}{*}{\cellcolor{gray!15}\textbf{GHOST}}& \multirow{-3}{*}{\cellcolor{gray!15}\textbf{BFT/Quorum}}& \multirow{-4}{*}{\cellcolor{gray!15}\textbf{Formalization}}\\ 
\arrayrulecolor{black}\hhline{---------}
\cellcolor{gray!15}
\textbf{Safety (consistency, persistence)}& \cite{gervais_security_2016,gazi_tight_2020,dembo_everything_2020,ramezan_analysis_2020}& \cite{garay_bitcoin_2015,gervais_security_2016,li_analysis_2019,ramezan_analysis_2020,yang_assessing_2020} & \cite{neu_availability-accountability_2022, sankagiri_blockchain_2021,buterin_incentives_2019, neu_ebb-and-flow_2021,kelkar_order-fair_2022,rana_optimal_2022} & \cite{neu_availability-accountability_2022, buterin_incentives_2019, wenting_securing_pos_2017,neu_ebb-and-flow_2021}& 
\cite{dembo_everything_2020,gazi_tight_2020,thomsen_formalizing_2021,kiayias_ouroboros_2017}& \cite{wang_byzantine_2022,neu_two_2022,neu_availability-accountability_2022,nakamura_refinement_2019,neu_ebb-and-flow_2021,li_formalizing_2020} &\cite{guerraoui_consensus_2019,bertrand_holistic_2022}&\multirow{6}{*}{\cite{graf_security_2021,badertscher_bitcoin_2017}}\\ 
\cline{1-8}
\cellcolor{gray!15}
\textbf{Liveness}& \cite{dembo_everything_2020,gazi_tight_2020, ramezan_analysis_2020}& \cite{garay_bitcoin_2015,li_analysis_2019,ramezan_analysis_2020}& 
\cite{sankagiri_blockchain_2021,buterin_incentives_2019,neu_availability-accountability_2022, neu_ebb-and-flow_2021,kelkar_order-fair_2022,rana_optimal_2022}& \cite{neu_availability-accountability_2022,buterin_incentives_2019, neu_ebb-and-flow_2021}& \cite{dembo_everything_2020,gazi_tight_2020,thomsen_formalizing_2021,kiayias_ouroboros_2017}& \cite{wang_byzantine_2022,neu_availability-accountability_2022,neu_two_2022, neu_ebb-and-flow_2021,nakamura_refinement_2019}&\cite{guerraoui_consensus_2019,bertrand_holistic_2022}& \\ 
\cline{1-8}
\cellcolor{gray!15}
\textbf{Common prefix, chain-quality, chain growth}& \cite{gervais_security_2016,gazi_tight_2020,dembo_everything_2020,ramezan_analysis_2020}& \cite{garay_bitcoin_2015,gervais_security_2016,dembo_everything_2020,ramezan_analysis_2020,li_analysis_2019}& \cite{sankagiri_blockchain_2021,neu_availability-accountability_2022, kelkar_order-fair_2022,rana_optimal_2022}& \cite{neu_availability-accountability_2022,neu_two_2022,neu_ebb-and-flow_2021}& \cite{dembo_everything_2020,gazi_tight_2020,thomsen_formalizing_2021,kiayias_ouroboros_2017}& \cite{neu_availability-accountability_2022,neu_two_2022,neu_ebb-and-flow_2021,li_formalizing_2020}& & \\ 
\cline{1-8}
\cellcolor{gray!15}
\textbf{Finality}&&&\cite{sankagiri_blockchain_2021,neu_availability-accountability_2022,buterin_incentives_2019,rana_optimal_2022,kelkar_order-fair_2022,neu_ebb-and-flow_2021}& \cite{neu_availability-accountability_2022,neu_two_2022,buterin_incentives_2019,neu_ebb-and-flow_2021}&& \cite{wang_byzantine_2022,neu_availability-accountability_2022,nakamura_refinement_2019,neu_ebb-and-flow_2021,li_formalizing_2020}&\cite{guerraoui_consensus_2019, bertrand_holistic_2022}& \\ 
\cline{1-8}
\cellcolor{gray!15}
\textbf{Dynamic availability}&& \cite{yu_unified_2020,garay_bitcoin_2015}&  \cite{sankagiri_blockchain_2021,neu_ebb-and-flow_2021,neu_availability-accountability_2022,rana_optimal_2022,kelkar_order-fair_2022}& \cite{neu_availability-accountability_2022,neu_two_2022,neu_ebb-and-flow_2021}& \cite{kiayias_ouroboros_2017}&\cite{neu_availability-accountability_2022,neu_ebb-and-flow_2021,neu_two_2022}&& \\ 
\cline{1-8}
\cellcolor{gray!15}
\textbf{"Other" formal properties} &
\cite{gervais_security_2016}$^{1}$, 
\cite{ramezan_analysis_2020}$^{2}$,
\cite{walck_tendrilstaller_2019}$^{6}$ & 
\cite{gervais_security_2016}$^{1}$, 
\cite{lee_countering_2019}$^{2}$,  
\cite{ramezan_analysis_2020}$^{2}$, 
\cite{yang_assessing_2020}$^{1,2}$, 
\cite{yu_unified_2020}$^{2}$ & 
\cite{kelkar_order-fair_2022}$^{5}$,
\cite{neu_availability-accountability_2022}$^{3}$,
\cite{rana_optimal_2022}$^{6}$,
\cite{sankagiri_blockchain_2021}$^{4}$&
\cite{neu_availability-accountability_2022}$^{3}$&  
\cite{yu_unified_2020}$^{2}$ &
\cite{neu_availability-accountability_2022}$^{3}$ &
\cite{bertrand_holistic_2022}$^{7}$ & \\
\arrayrulecolor{black}\hhline{---------}
\end{tabular}%
}
\small
\begin{tabular}{llll}
$^1$ Block size, network delay, stale blocks rate. &
$^2$ Reward distribution. &
$^3$ Availability-accountability dilemma. \\
$^4$ Availability-finality dilemma. 
$^5$ Order-fairness. & 
$^6$ Bootstrapping. &
$^7$ System convergence. &
\end{tabular}

\end{table*}

\subsection{Blockchain security formalization}
\label{sec:formal}
Blockchain formalization and abstraction are fundamental in characterizing the architectural model of blockchain to prove security properties. In their seminal work,~\citet{garay_bitcoin_2015} provide the first formal proof and formalize blockchain security properties under synchronous network conditions drawing from the longest-chain fork-choice rule design (i.e., everyone agrees to append blocks to the longest chain seen, weighted by difficulty) and the~\ac{POW} Sybil resistance mechanism. The authors introduce three essential security properties that represent necessary conditions to guarantee safety and liveness: 1)~\textit{common prefix} describes the existence of a large commonly agreed sub-chain and provides probabilistic safety guarantees, 2)~\textit{chain quality} describes the ratio at which honest blocks are included in the chain and represents a liveness property, and 3)~\textit{chain growth}, the speed at which the chain grows, i.e., keeps recording blocks.\\[-0.35cm]

\textbf{Common prefix} with security parameter $k\in \mathbb{N}$~\cite{garay_bitcoin_2015, kiayias_ouroboros_2017}: For any node~$l$, let $\mathcal C_l$ the current view of the chain and $\smash{\mathcal C_l^{\lceil{k}}}$ denote this chain with the last $k$ blocks removed. Then any two honest nodes $i$ and $j$ have consistent view of the chain up to the $k$ last blocks, i.e.,  $\smash{\mathcal{C}^{\lceil{k}}_{i} \preceq \mathcal{C}^{\lceil{k}}_{j}}$ or $\smash{\mathcal{C}^{\lceil{k}}_{j} \preceq \mathcal{C}^{\lceil{k}}_{i}}$, where $\preceq$ denotes ``is prefix of''. \\[-0.35cm]

\textbf{Chain-quality} with parameter $\mu \in [0,1]$~\cite{kiayias_ouroboros_2017}: For any reported chain from the common prefix of an honest node, the ratio of adversarial blocks is at most $1-\mu$. \\[-0.35cm]

\textbf{Chain-growth} with parameter $\tau \in [0,1] $~\cite{garay_bitcoin_2015,kiayias_ouroboros_2017,neu_availability-accountability_2022}: Let $\mathcal{C}_t$ be the chain at time t for $t\geq 0$ and $\lvert \mathcal C_t\rvert$ the length of $\mathcal C_t$ . Then $\mathcal{C}$ has chain growth $\tau$ if $\lvert\mathcal{C}_t\rvert - \lvert \mathcal{C}_s \rvert \geq \tau \cdot (t-s)$ for all $0\leq s\leq t$.

The formalization and blockchain security notions in~\cite{garay_bitcoin_2015} are also leveraged by many other works for analyzing the formal security aspects of different blockchain consensus designs~\citep[e.g.,][]{li_analysis_2019, sankagiri_blockchain_2021, kelkar_order-fair_2022, neu_availability-accountability_2022,gervais_security_2016, dembo_everything_2020, thomsen_formalizing_2021, rana_optimal_2022, ramezan_analysis_2020}. 
\citep{badertscher_bitcoin_2017} complements previous work by constructing a UC-secure~\ac{POW} longest chain blockchain under synchrony. The UC framework~\cite{canetti2007universally} serves as an abstract and general model to define composable protocol functionalities, providing strong security guarantees also in concurrent protocol execution. The ideal functionality in~\citep{badertscher_bitcoin_2017} enables the analysis of various customizable properties, including synchrony assumptions and adversarial capabilities.~\cite{graf_security_2021} generalize the ideal functionality of blockchains, including~\citep{badertscher_bitcoin_2017}, in the form of an ideal ledger that reflects the broad spectrum of blockchain properties (e.g., consensus mechanism and synchronization models). Their ideal ledger comprises a globally ordered list of transactions on a global state, which can be interacted with through various actionable subroutines (e.g., read/write operations). Alternative forms of abstraction, such as automata, are also suitable to prove safety and liveness properties~\cite{bertrand_holistic_2022}. To further evaluate security notions,~\cite{li_analysis_2019} strengthens previously defined security properties, namely \textit{common prefix}, \textit{chain quality}, \textit{safety}, \textit{liveness}, and \textit{chain growth}~\cite{garay_bitcoin_2015}.

\Ac{POS} addresses safety against long-range attacks by offering finality guarantees (either immediate, through check-pointing~\cite{buterin_incentives_2019}, or probabilistic~\cite{kiayias_ouroboros_2017}). As the current total stake is a publicly visible figure~\cite{pye_resource_2020}, they must involve a \ac{BFT}-style routine with ``honest supermajority'' assumption for immediate finality or ``honest majority'' for probabilistic finality guarantees. 
Consequently, they cannot tolerate an unknown threshold of stake represented by inactive or disconnected participants~\cite{lewispye2021}. On the other hand, the total hash rate of a \ac{POW}-based permissionless blockchain system is unknown and can fluctuate unpredictably with nodes joining, leaving, or adapting their mining efforts. Therefore, prioritizing dynamic availability seems natural for longest-chain \ac{POW} constructions. 

One property of blockchains that prioritize safety is to guarantee that no two different blocks at the same height are finalized. In general, \ac{BFT} protocols such as Casper identify adversarial behavior and forfeit their stake in the event of equivocation as long as the majority stake is controlled by honest nodes~\cite{neu_availability-accountability_2022}. Generally, the nodes in control of the majority stake would be in a position to signal adversarial nodes' behavior and hold them accountable for their acts. However,~\cite{neu_availability-accountability_2022} shows conflicts between dynamic availability and accountability with respect to safety and liveness, thus defining another dilemma. On the one hand, permissioned~\ac{BFT} protocols provide safety and can tolerate up to one-third of adversarial participants~\cite{bertrand_holistic_2022,neu_availability-accountability_2022} in non-synchronous networks while remaining accountable, as participants are identified within the network~\cite{neu_availability-accountability_2022}. In contrast, dynamically available blockchains provide liveness regardless of the specific Sybil resistance mechanism~\cite{neu_availability-accountability_2022}, with the constraint that the majority of the resource must be controlled by honest nodes~\cite{garay_bitcoin_2015,li_analysis_2019,dembo_everything_2020}. Common \ac{POS} gadgets such as Casper, and Gasper satisfy safety and provide finality and accountability but lack strong liveness guarantees, as demonstrated by known attacks~\cite{neu_availability-accountability_2022, neu_two_2022,nakamura_refinement_2019,li_formalizing_2020}. More details on~\ac{POS} gadgets can be found in \Cref{sec:pos}.

As the dynamic availability-finality dilemma and the dynamic availability-accountability dilemma provably cannot be resolved with a single consensus design, hybrid solutions utilizing dual ledgers emerge as a viable approach to tackle this trade-off. Dual ledgers leverage two different, ``user-dependent'' sets of consensus rules. For instance, while one main chain is dynamically available, tolerating nodes leaving and joining, the other is a check-pointed prefix of the previous consisting of finalized blocks, thus prioritizing safety in the event of de-synchronization~\cite{sankagiri_blockchain_2021,neu_availability-accountability_2022}. \cite{neu_availability-accountability_2022} shows that existing accountable and safe~\ac{BFT} protocols can also be used all together as part of the consensus rules pursuing dual ledger strategies~\cite{neu_availability-accountability_2022,neu_ebb-and-flow_2021}. 

\subsection{Bootstrapping and blockchain continuity}\label{sec:boots}
Permissionless blockchains' inherent properties are not limited to security aspects regarding continuous operation but also the seamless process of synchronizing nodes. Any de-synchronized node -- whether because of a newly spawned node or temporarily in a partition -- must be able to obtain a verifiable latest state of the protocol, i.e., to reach the subset of blocks that form the common prefix. Bootstrapping is the process by which nodes synchronize their local state with the globally agreed state~\cite{rana_optimal_2022}.
The ability of a blockchain to allow such verifiable bootstrapping with minimal trust assumptions is that the blockchain is objective~\cite{vashchuk_pros_2018}. One powerful notion of permissionless \ac{POW} blockchains with the longest-chain rule weighted by difficulty is that they are objective~\cite{vashchuk_pros_2018}.
Nodes can reliably reach the latest chain state by locally reconstructing and verifying the chain without any external contribution as long as the node is connected to at least one other honest node. Note that all permissionless blockchains are required to agree on an initial trusted source for the genesis block and node software as public given parameters. In contrast, \ac{POS} blockchains are weakly subjective: Nodes need external sources of information such as an additional set of recent blocks agreed to be valid (i.e., check-pointed) to determine which is the latest agreed state and, thus, to identify the canonical chain~\cite{vashchuk_pros_2018}.

Bootstrapping a blockchain is an important characteristic that can be negatively affected by eclipse attacks~\cite{walck_tendrilstaller_2019} or forking events, e.g., in long-range attacks in~\ac{POS}~\cite{rana_optimal_2022, deirmentzoglou_survey_2019}. In eclipse attacks, adversarial nodes supply invalid blocks to their neighboring nodes, therefore disrupting the synchronization process~\cite{rana_optimal_2022}. Check-pointing methods as suggested in~\citep{sankagiri_blockchain_2021} can address this issue~\cite{rana_optimal_2022}. Check-pointing and, thus, relying on an oracle to query external messages for validation is a requirement for safety in~\ac{POS} protocols~\cite{lewispye2021}. 
Therefore, there seems to be a correspondence in blockchain design, where objective blockchains (e.g.,~\ac{POW}) only provide probabilistic immutability, in other words, probabilistic finality, whereas weakly subjective ones such as~\ac{POS} can achieve finality.
\subsection{Formal security analysis of PoW}\label{sec:pow}
Going beyond the original heuristics presented in the Bitcoin Whitepaper~\cite{nakamoto_bitcoin_2009},~\cite{garay_bitcoin_2015} formally proves that safety (i.e., \textit{common prefix}) is satisfied with high probability only under the 
assumption of honest majority ($>50\,\%$, weighted by hash rate), and a tightly connected network, i.e., message delivery of blocks is fast compared to the block production rate, which represents a stronger assumption than synchrony. \cite{garay_bitcoin_2015} also proves that there is a maximum number of blocks, and waiting time after which an honest transaction is guaranteed to be included, thus defining the liveness (i.e., \textit{chain quality}) property
~\cite{yang_assessing_2020}. \cite{gazi_tight_2020}, which builds upon~\cite{garay_bitcoin_2015}, then proves the common prefix and the liveness property including \textit{chain growth} under partial synchrony.

In \ac{POW}, a node's share of total invested computational power in the blockchain network is equal to the probability at which a new block building on a given latest block can be found by either party (both for honest and adversarial participants)~\cite{garay_bitcoin_2015}. However, the provable \textit{chain quality} starts to degrade significantly when the adversarial threshold approaches $50\,\%$, suggesting a potentially asymmetric increase in the share of blocks proposed in comparison to the share of computational power~\cite{garay_bitcoin_2015}. Indeed, \cite{eyal_18_selfish} shows that by adopting ``selfish mining'' where an entity mines on a private chain and strategically delays the release of blocks, adversarial nodes can gain a disproportionate advantage (i.e., contribute more blocks to the canonical chain) and, therefore, also earn more rewards than honest nodes~\cite{yang_assessing_2020, gervais_security_2016}. As such, \ac{POW} longest-chain protocols fail to appropriately reward participation according to their portion of the share of the hash rate~\cite{garay_bitcoin_2015}. The probability of successfully publishing a private chain, thus inducing a reorganization of blocks, still increases exponentially with the adversary's share of voting power~\cite{ramezan_analysis_2020, gervais_security_2016}. More precisely, the relative revenue of selfish mining depends on the fork-choice rule followed by honest nodes~\cite{yang_assessing_2020} and other factors such as the stale block rate (i.e., valid blocks that collide with others resulting in non-inclusion) and network partitions, i.e., forks~\cite{yang_assessing_2020,gervais_security_2016}. These findings show the need for tighter upper bounds on the adversarial threshold of computational power given the side effects of message delays, and the throughput of block creation to guarantee the common prefix and chain quality~\cite{garay_bitcoin_2015}. Simulation results in~\cite{yang_assessing_2020} confirm these upper bounds. As the share of the hash rate increases beyond~33\,\%, the relative revenue of selfish mining increases disproportionally~\cite{gervais_security_2016, yang_assessing_2020}. On the other hand, the capacity of successfully and selfishly mining multiple consecutive blocks, thus causing a reorganization, still decreases exponentially with each new included block in the canonical chain~\cite{nakamoto_bitcoin_2009,garay_bitcoin_2015, li_analysis_2019}. Reorganizations benefit adversaries either by earning additional block rewards (degrading the chain quality) or as a consequence of completing successful double spends (compromising safety)~\cite{garay_bitcoin_2015}. \cite{dembo_everything_2020}~shows that selfish mining is indeed the worst possible adversarial attack on the longest-chain-based \ac{POW} consensus. Incorporating random choice in the fork-choice rule for a node that learns about two different longest (in particular, equally long) chains at roughly the same time mitigates the ``worst-case'' achievable through selfish-mining~\cite{eyal_18_selfish}, with a corresponding upper bound of $\approx 33\,\%$ on the adversarial threshold.

Other factors, such as network latency and block propagation time, also influence the security of \ac{POW}-based consensus protocols~\cite{gervais_security_2016}.~\cite{dembo_everything_2020} demonstrates that the probability of adversarial events is a function of the block propagation rate and the adversary's share of the computational power. A slow block production rate offers higher levels of security~\cite{ramezan_analysis_2020} as it benefits consistency~\cite{gervais_security_2016}, ensures the common prefix security property~\cite{garay_bitcoin_2015}, and keeps the probability of successful double-spends low~\cite{ramezan_analysis_2020}. \cite{gervais_security_2016} also shows that lowering the propagation rate of valid block inclusion up to a certain threshold that depends on the average propagation time (e.g., approx.~1~minute in Bitcoin) does not considerably affect the security assumptions.
\subsection{Formal security analysis of PoS}\label{sec:pos}
\Ac{POS} protocols have been proposed and implemented similarly to the original~\ac{POW} in Bitcoin by heuristically assuming its security as in~\citep{king_ppcoin_2012}. Several attacks that severely affect the common prefix of the ledger need to be accounted for in~\ac{POS} blockchains~\citep{wenting_securing_pos_2017}; most prominently, \textit{nothing-at-stake} and ~\textit{long-range} attacks (see \Cref{sec:2-block}). Additionally,~\ac{POS} can be vulnerable to~\textit{grinding} attacks, where nodes exploit weaknesses of pseudo-random number generators used for electing block proposers in order to gain an advantage in the probability of being elected~\citep{kiayias_ouroboros_2017, deirmentzoglou_survey_2019}. \citep{wenting_securing_pos_2017} shows two ways of tackling such attacks: by authenticating and binding nodes' identities to proposed blocks, and by using trusted hardware for block production. \citep{kiayias_ouroboros_2017} presented the first formalization for a~\ac{POS} blockchain and proved~\textit{consistency} and~\textit{liveness} guarantees with respect to an adversarial threshold close to $50\,\%$.

Subsequent works on~\ac{POS} constructions have been centered around mitigating or preventing these attacks to guarantee the security properties of blockchain such as safety (\textit{consistency} and \textit{common prefix}) and liveness (\textit{chain-quality} and \textit{chain growth}). \ac{POS} in synchronous networks with the longest-chain rule, when assuming that the election process of block proposers is a random process~\cite{kiayias_ouroboros_2017}, maintains similar security properties to~\ac{POW} with respect to safety and liveness, including~\textit{chain quality},~\textit{chain growth}, and~\textit{common prefix}~\cite{gazi_tight_2020, dembo_everything_2020,thomsen_formalizing_2021}. The key difference lies in the honest majority requirement, with~\ac{POS} necessitating $>66\,\%$ of the stake to be controlled by honest and participating peers~\cite{thomsen_formalizing_2021, nakamura_refinement_2019} compared to $>50\,\%$ in~\ac{POW}~\cite{garay_bitcoin_2015}. \ac{POS} necessitates the depositing of a stake to an intrinsic \ac{PKI}. The stake is relative to the known number of public keys in the system, thereby heuristically fostering a BFT protocol. Further, its dynamic availability can be considered analogous to a non-synchronous network. Interpreting the underlying protocol as~\ac{BFT}-style, consensus can be achieved if at least $\approx 66\,\%$ of the voting power is controlled by honest nodes. Alternatively, consensus protocols that integrate several key components such as a new rewarding scheme, a modified longest-chain fork-choice rule, and forward-secure signatures can result in an approximate Nash equilibrium that disincentivizes validators from deviating from the protocol and thus reduce the tolerable adversarial threshold to $<50\,\%$~\cite{kiayias_ouroboros_2017}. Note that these schemes differ substantially from the ones coined from the \ac{POS} variants proposed by Peercoin~\cite{king_ppcoin_2012} and implemented in Ethereum~\cite{neu_ebb-and-flow_2021}. In particular, \cite{kiayias_ouroboros_2017} incorporates a publicly verifiable random function that ensures the generation of a globally verifiable random value, effectively tackling grinding attacks.

\cite{buterin_incentives_2019} introduce some modifications to the~\ac{POS}-based longest-chain protocols in the form of Casper (see \Cref{sec:2-block}) to address long-range attacks. Casper finalizes blocks of the canonical chain and holds adversarial nodes accountable (e.g., through slashing), mitigating the nothing-at-stake problem~\cite{buterin_incentives_2019, neu_availability-accountability_2022}. \cite{nakamura_refinement_2019} and~\cite{li_formalizing_2020} present a set of formal proofs, grounded in prior reasoned arguments by \cite{buterin_incentives_2019} on the safety of Gasper. The authors show safety for Gasper for up to $33\,\%$ of adversarial stake. As Gasper employs the GHOST protocol, orphaned/stale blocks that are valid and received votes still influence the chain~\cite{yang_assessing_2020,neu_two_2022}. These events influence the security of the system by facilitating new attacks that may harm the liveness of the protocol~\cite{yang_assessing_2020}. Indeed, 
~\cite{wang_byzantine_2022,neu_two_2022} show a lack of liveness of this implementation. \cite{neu_two_2022} further finds vulnerabilities in two variants of the GHOST protocol that can be exploited with even less than $33\,\%$ of adversarial stake. Both designs suffer from variants of long-range attacks and equivocation in combination with leveraging the influence of orphaned blocks to displace or split the canonical chain. Owing to costless simulation, nodes can generate several blocks in Gasper and equivocally vote for them, which potentially leads to an ``avalanche" or ``balancing'' attack. As a result, the canonical chain can be displaced or forked for an undefined amount of time~\cite{neu_two_2022,neu_ebb-and-flow_2021}. Displacement shifts grow the chain horizontally and vertically by leveraging the voting weights of equivocating blocks. Withheld blocks are altogether released horizontally on top of the same privately mined block (i.e., in the form of an ``avalanche''), thus shifting (i.e., forking) the canonical chain. This results in a lack of safety and liveness of the protocol~\cite{neu_two_2022}. To address this issue, a modification called Latest Message Driven~(LMD) was introduced~\cite{neu_two_2022}. The LMD~modification affects the decision on GHOST by accounting only for the latest voted block instead of an unbounded number of previous blocks so that it tackles equivocating on multiple equal height blocks~\cite{neu_ebb-and-flow_2021}. However,~\cite{neu_two_2022} and~\cite{neu_ebb-and-flow_2021} describe a ``balancing" attack that affects the LMD~GHOST implementation, thus demonstrating a potential lack of safety as a consequence of an undefined but constant chain split. An adversary with a small fraction of the stake can timely equivocate on blocks to make both chains grow at the same time. Notably, only the first attempt at forking the chain view by equivocating represents a slashable action. Both~\cite{wang_byzantine_2022} and~\cite{neu_two_2022} describe scenarios where in the event of a chain fork, an adversary is able to keep a split view of the chain such that half of the \textit{validators} see either side. As a consequence, no chain is ever finalized, thus breaking the liveness property. 
\section{Discussion and Conclusion}
This paper focuses on comparing the formal security aspects of~\ac{POW}- and~\ac{POS}-based consensus mechanisms. We answer \textit{RQ1} by consolidating established security notions and corresponding distributed computing impossibility results.
The formal blockchain security properties we identified are \textit{safety}, \textit{consistency}, \textit{common prefix}, \textit{finality}, \textit{liveness},  \textit{chain quality}, \textit{chain growth}, and \textit{dynamic availability}. Additionally, the paper highlights that \textit{safety} is related to the \textit{common prefix}, \textit{consistency}, and \textit{finality} properties. We discuss impossibility results including FLP~\cite{fischer_impossibility_1985}, CAP~\cite{brewer_cap_theorem,gilbert_lynch_web_02} 
to emphasize key theoretical limitations of distributed networks and outline ``upper bounds'' on security properties of \ac{POW}- and \ac{POS}-based permissionless blockchains. We also point out that these limitations can, to some extent, be overcome by defining security properties probabilistically with respect to a security parameter~\cite{fischer_impossibility_1985, pye_resource_2020,lewispye2021} (e.g., probabilistic finality instead of absolute finality~\cite{pye_resource_2020}) and by strengthening assumptions about network partitions, i.e., synchrony properties~\cite{pye_resource_2020}. Due to the discussed impossibility results and dilemmas, modeling security in permissionless networks needs relatively strong assumptions on network synchrony to assure safety and liveness~\cite{neu_availability-accountability_2022,sankagiri_blockchain_2021,lewispye2023byzantine}. 
Furthermore, non-deterministic protocols are the only known approaches to solving consensus in permissionless networks with inherently dynamic voting power distribution~\cite{lewispye2023byzantine}. \cite{wang_byzantine_2022}~notes some evaluation disparity between academic and practical design implementations of consensus protocols concerning security properties. While the former often reason based on tight assumptions to prove security properties rigorously, the latter tend to relax the adversarial capabilities to accommodate the desired proofs.

Our path toward answering RQ2 involves several key observations. For both~\ac{POW} and~\ac{POS}-based consensus with the longest-chain rule, under specific honest majority assumptions after some reasonable time period, blocks are part of \textit{common prefix}~\cite{garay_bitcoin_2015, neu_two_2022}. In general, blocks are included in the canonical chain in~\ac{POW} with overwhelming probability (i.e., \textit{probabilistic finality}). Similar patterns can be observed in probabilistic~\ac{POS} constructions such as~\cite{kiayias_ouroboros_2017}, while finality is immediate or satisfied after a certain finite time in~\ac{POS} constructions using~\ac{BFT} protocols. Every honest node ends up sharing a common subset of blocks (i.e., \textit{consistency})~\cite{garay_bitcoin_2015, neu_two_2022, dembo_everything_2020}. \Ac{POW} and probabilistic~\ac{POS} variants achieve these properties if honest nodes control the majority of the voting power ($>50\,\%$) when assuming tight and favorable network conditions. \ac{BFT}-based~\ac{POS}, on the other hand, requires honest nodes to control over $66\,\%$ of the voting power. However, \ac{POW} also requires an honest $66\,\%$ majority to ensure sufficient chain growth and quality due to the profitability of selfish mining with a voting power above $33\,\%$~\cite{garay_bitcoin_2015, yang_assessing_2020}. \ac{POS} longest-chain protocols can achieve comparable thresholds in networks where synchronicity is guaranteed~\cite{thomsen_formalizing_2021,kiayias_ouroboros_2017}. \ac{POS} constructions that depend on accountable gadgets such as Casper can guarantee safety regardless of the fork-choice rule even under powerful adversaries~\cite{wang_byzantine_2022}. Liveness was initially posited to be guaranteed in a static validator setting and a supermajority of honest controllable voting power~\cite{nakamura_refinement_2019} and later proven under tightly synchronized networks assuming $<50\,\%$ of adversarial controlled stake in~\cite{thomsen_formalizing_2021}. 

We also found some inherent trade-offs affecting both \ac{POW} and \ac{POS}, such as the \textit{dynamic availability-finality} dilemma~\cite{lewispye2021, lewispye2023byzantine, sankagiri_blockchain_2021} and the \textit{availability-accountability}~\cite{neu_availability-accountability_2022} dilemma. Balancing these trade-offs involves prioritizing specific designs based on the desired capabilities of the network. \ac{POS} with dual ledgers are placed in as a potential solution to overcome such trade-offs between safety and liveness. Particularly, Gasper provides liveness under a simple honest majority assumption ($>50\,\%$) and satisfies the \textit{common prefix} property if a supermajority ($>66\,\%$) of stake is honest~\cite{neu_two_2022}. Similar thresholds have been shown for other accountable and safe~\ac{BFT} protocols~\cite{neu_availability-accountability_2022,neu_ebb-and-flow_2021}.

Researchers also incorporate slightly less common formal security or other blockchain properties, such as the \textit{network delay} on message communication,  the \textit{stale block ratio}, and the \textit{order-fairness} of transactions~\cite{kelkar_order-fair_2022}. Our review indicates that if the block propagation rate is sufficiently slow to ensure a low stale block ratio, \ac{POW}-based Nakamoto consensus offers stronger formal security guarantees regarding \textit{common prefix}, \textit{consistency}, and \textit{liveness}~\cite{gervais_security_2016, yang_assessing_2020} than \ac{POS}. This property can be traced back to \ac{POW}'s choice of the utilization of computational resources in achieving Sybil resistance, as investing computational resources is inherently dynamic in nature. \Ac{POW} causes substantial costs for block production that inherently avoids the need to address equivocation and long-range attacks. Nevertheless, longest-chain consensus variants with~\ac{POS}-based construction provide similar guarantees when specific attacks are addressed~\cite{buterin_incentives_2019, thomsen_formalizing_2021}. On the other hand, while~\ac{POS} with GHOST allows for higher block production rates and, therefore, lower latencies and higher throughput when storage is not the bottleneck~\cite{gervais_security_2016,yang_assessing_2020}, it comes at the expense of increased complexity and weaker formal security guarantees~\cite{neu_two_2022}. \Ac{POS} gadgets aim to ensure \textit{safety} by providing \textit{finality}, but can be compromised with non-zero probability with a small fraction of adversarially controlled stake~\cite{neu_two_2022}. 

According to our~\ac{SLR}, no single solution currently addresses all of the desired security properties. It seems that~\ac{POW} and~\ac{POS}-based consensus protocols have already approached their optimal design, given the constraints posed by the impossibility results and dilemmas surveyed. Consequently, there is a line of work on gadgets that aim to satisfy various blockchain security properties in the form of dual ledgers to circumvent the impossibility results and dilemmas~\cite{neu_ebb-and-flow_2021, neu_availability-accountability_2022, sankagiri_blockchain_2021}. Each ledger prioritizes one security property, e.g., \textit{safety} during partitions (i.e., finalized blocks) or liveness (i.e., the chain keeps growing and is dynamically available), and each user can pick their priority depending on the intended transaction.

We emphasize that in practice, formal security guarantees should not only be reduced to the adversarial hash power or stake threshold. To give a recent example, a bug in a very common Ethereum client implementation recently caused the check-pointed chain to stop finalizing blocks for around an hour~\cite{nijkerk_2023}, thus inducing a state where liveness was not guaranteed in the safety-prioritizing chain. Nevertheless, as a consequence of the dual-ledger constructions, the dynamically available chain kept growing with transactions being included, so Ethereum did not suffer a full liveness issue~\cite{neu_availability-accountability_2022}. After the issue was resolved the network recovered to its normal state, indicating a high degree of resilience~\cite{neu_availability-accountability_2022}. This incident suggests that analyzing the levels of decentralization and setting them in relationship with the tolerable faulty thresholds in the employed consensus mechanism is not only critical in the distribution of computational power or stake but also on other layers~\cite{sai2021taxonomy}. Finally, consensus designs for permissionless blockchains do not only impact formal security properties but account for and balance also other important issues, such as economic security aspects (including long-term (de-)centralization tendencies) and performance. A broader consideration of~\ac{POW} and~\ac{POS} is, therefore, required for a holistic perspective of security in permissionless blockchain designs.

\begin{acks}
    This research was funded in part by the Luxembourg National Research Fund (FNR) through the PABLO project (grant reference 16326754), the FiReSpARX Project (grant reference 14783405), by PayPal, grant reference ``P17/IS/13342933/PayPal-FNR/Chair in DFS/Gilbert Fridgen'' (PEARL), and by the Bavarian Ministry of Economic Affairs, Regional Development and Energy through the project ``Fraunhofer Blockchain Center (20-3066-2-6-14)''. For the purpose of open access, and in fulfillment of the obligations arising from the FNR grant agreement, the authors have applied a Creative Commons Attribution~4.0 International (CC~BY~4.0) license to any Author Accepted Manuscript version arising from this submission. We also thank Joaquín D. Fernández, Orestis Papageorgiou, and Nadia Pocher for their valuable feedback. 
\end{acks}

\bibliographystyle{ACM-Reference-Format}
\bibliography{ref} 

\end{document}